# BPF-type Region-of-interest Reconstruction for Parallel Translational Computed Tomography

Weiwen Wu[a], Hengyong Yu[b], Shaoyu Wang[a], Fenglin Liu[a,c,*]

[a]*Key Lab of Optoelectronic Technology and Systems, Ministry of Education, Chongqing University, Chongqing, China*

[b]*Department of Electrical and Computer Engineering, University of Massachusetts Lowell, Lowell, MA, USA*

[c]*Engineering Research Center of Industrial Computed Tomography Nondestructive Testing, Ministry of Education, Chongqing University, Chongqing, China*

**Abstract.** Recently, an ultra-low-cost linear scan based tomography architecture was proposed by our team. Similar to linear tomosynthesis, the source and detector are translated in opposite directions and the data acquisition system targets on a region-of-interest (ROI) to acquire data for image reconstruction. This kind of tomography architecture was named parallel translational computed tomography (PTCT). In our previous studies, filtered backprojection (FBP)-type algorithms were developed to reconstruct images from PTCT. However, the reconstructed ROI images from truncated projections have severe truncation artifacts. In this paper, we propose two backprojection filtering (BPF)-type algorithms named MP-BPF and MZ-BPF to reconstruct ROI images from truncated PTCT data. A weight function is constructed to deal with data redundancy for multi-linear translations modes. Extensive numerical simulations are performed to evaluate the proposed MP-BPF and MZ-BPF algorithms for PTCT in fan-beam geometry. Qualitative and quantitative results demonstrate that the proposed BPF-type algorithms can not only accurately reconstruct ROI images from truncated projections but also provide high-quality images for the entire image

*Corresponding author: Fenglin Liu, Key Lab of Optoelectronic Technology and Systems, Ministry of Education, Chongqing University, Chongqing 400044, China. Tel.: +86 23 8639 4828; E-mail: liufl@cqu.edu.cn

support in some circumstances.

## 1. Introduction

A new low-end computed tomography (CT) architecture was recently proposed for developing countries [1]. Because its X-ray source and detector are parallel translated in opposite direction, we named it as parallel translational computed tomography (PTCT). Filtered backprojection (FBP)-type algorithms were successfully developed for PTCT [2]. They can accurately reconstruct images from complete or nontrucated PTCT projections. However, it is very common that the detector can only cover part of the object resulting in incomplete and truncated projections. Those truncated projections require more complicated reconstruction algorithms than complete or nontruncated projections, and it is impossible to perform theoretically exact image reconstruction in some circumstances [3, 4]. On the other hand, it is of practical significance to develop algorithms to accurately reconstruct region-of-interest (ROI) from truncated projections collected on PTCT [5, 6]. Unfortunately, such algorithms are not available yet. This motivates us to investigate and develop advanced algorithms to reconstruct ROI images from PTCT truncated projections.

More than one decade ago, backprojection filtration (BPF)-type algorithms, based on the concept of PI-line or chord, were proposed for ROI reconstruction from truncated projections in general scanning configurations [2, 4, 7-13]. BPF-type algorithms can be divided into three steps for helical/spiral scan [14, 15]: 1) the differentiation of projection data; 2) weighted backprojection; 3) finite inverse Hilbert filtering along PI-line segments. To approximately reconstruct off-midplane images from circular scanning trajectory, similar to the classic FDK algorithm, the virtual source locus and virtual PI-lines were also introduced [5]. However, there are some scanning configurations that do not have PI-lines or virtual PI-lines, for example, single or multiple straight-line scanning trajectory [16, 17]. Indeed, the applications of straight-lines locus or multi-line-segment locus lie in linear tomosynthesis [18, 19], in security, industrial, or business scanning, such as luggage inspection [20] or wood identification [21], and potentially even for the hexagonal bar PET [22].

In this paper, we will develop BPF-type algorithms for PTCT. To reconstruct high quality images from straight-line scanning configuration using BPF-type algorithms, the concept of PI-lines and PI-line segments should be modified. A line parallel to straight-line scanning locus and passing through the object support is chosen as linear-PI-line (L-PI). Furthermore, to deal with data redundancy for multiple linear translations scanning trajectories, a normalized weight function will be constructed. Another key step of BPF-type algorithms is the finite inverse Hilbert transform [23]. The constant C in the finite inverse formulae from [4, 10] needs to be determined directly from measured projections. However, in PTCT, we cannot obtain this specific value of object function because any ray emitted from x-ray source is impossible parallel to L-PI lines. Therefore, we adopt two explicit formulae reported in [24-26] and [27], which can be able to eliminate the effect of constant in the finite inverse Hilbert transform. As a result, we obtain two BPF-type algorithms which are named MZ-BPF and MP-BPF algorithm, respectively.

The rest of this paper is organized as follows. In section II, we present generalized PTCT model and review FBP-type algorithm. In section III, based on the concept of L-PI line, we derive MZ-BPF and MP-BPF algorithms for PTCT. In section IV, we show the results of numerical studies from nontruncated and truncated projections. In section V, we discuss some related issues and conclude the paper.

**2. Review of PTCT FBP Algorithm**

We first briefly review the PTCT geometry [1, 2, 28]. In PTCT, the imaging object is stationary and the x-ray source and detector are translated in opposite directions (Fig.1). Assuming a global coordinate system attached to the imaging object, the source locus can be expressed as

$$\mathbf{\Phi}(\psi,\lambda) = \left( \frac{-h}{\cos(\psi+\lambda)}\sin\lambda, \frac{-h}{\cos(\psi+\lambda)}\cos\lambda \right), \text{ s.t } \lambda \in \left(-\frac{\pi}{2}-\psi, \frac{\pi}{2}-\psi\right), \quad (1)$$

where $\lambda$ denotes the angle between the vector from x-ray source to origin and $y$ axis, $h$ and $d$ are the distances from the x-ray source trajectory to the origin and detector trajectory, and $\psi$ is the angle between scanning trajectory and $x$ axis. In this work, we assume a compactly supported object function $f(\mathbf{r})$ where $\mathbf{r} := (x,y)$ is a 2D position vector.

Now, we introduce a moving local coordinate system whose origin is the source point. In the fixed global coordinate system, two unit vectors of the moving coordinate system are defined as

$$\begin{aligned} \mathbf{d}_1 &= (-\sin\psi, \cos\psi) \\ \mathbf{d}_2 &= (\cos\psi, \sin\psi) \end{aligned}. \quad (2)$$

In this paper, we adopt a flat panel detector for PTCT [1, 2] and only consider the fan-beam geometry for the central slice. Any element on the detector can be indexed by a parameter $t$, and $t=0$ corresponds to the x-ray passing through the global coordinate origin. The projection $p(t,\lambda,\psi)$ is the line integral along the x-ray path emitting from the x-ray source $\mathbf{\Phi}(\psi,\lambda)$ to the detector element at position $t$. Thus, the projection $p(t,\lambda,\psi)$ of an object function obtained by the detector at $t$ can be written as

$$p(t,\lambda,\psi) = \int_{-\infty}^{+\infty} f\left(\mathbf{\Phi}(\psi,\lambda)+l\mathbf{e}\right)dl, \qquad (3)$$

where the unit vector $\mathbf{e}$ is x-ray direction starting from $\mathbf{\Phi}(\psi,\lambda)$ to $\mathbf{r'} \in R^2$ on the object image

$$\mathbf{e} = \frac{\mathbf{r'}-\mathbf{\Phi}(\lambda,\psi)}{|\mathbf{r'}-\mathbf{\Phi}(\lambda,\psi)|}. \qquad (4)$$

and

$$t = d\left(\frac{\mathbf{e}\cdot\mathbf{d}_2}{\mathbf{e}\cdot\mathbf{d}_1} - \tan(\lambda+\psi)\right). \qquad (5)$$

Based on the aforementioned definitions, an approximate image can be reconstructed by an FBP algorithm as [28]

$$f(\mathbf{r}) = \int_{\lambda_b}^{\lambda_e} \frac{d^2}{U^2} \int_{t_{\min}}^{t_{\max}} \frac{h\sec(\lambda+\psi)}{\sqrt{d^2+2td\sin(\lambda+\psi)+t^2}} p(t,\lambda,\psi) g(t'-t)\,d\lambda dt, \qquad (6)$$

where $\lambda_b$ and $\lambda_e$ are the corresponding angles of the start and end positions of the x-ray source trajectory, $U = h + \mathbf{r}\cdot\mathbf{d}_1$ is the distance from reconstruction point to x-ray source scanning locus, $t'$ is the local coordinate for the x-ray passing through the point $\mathbf{r}$ and its value can be determined by using Eqs. (4) and (5) as

$$\mathbf{e} = \frac{\mathbf{r} - \boldsymbol{\Phi}(\lambda,\psi)}{|\mathbf{r} - \boldsymbol{\Phi}(\lambda,\psi)|} \quad \text{and} \quad t' = d\left(\frac{\mathbf{e}\cdot\mathbf{d}_2}{\mathbf{e}\cdot\mathbf{d}_1} - \tan(\lambda+\psi)\right), \text{ and } g(t) \text{ is the ramp-filter}$$

$$g(t) = \int_{-\infty}^{+\infty} |\rho| e^{2\pi i \rho t} d\rho. \tag{7}$$

It should be pointed out that Eq.(6) can provide theoretically exact reconstruction results if the scanning trajectory is from minus infinity to infinity. However, Eq.(6) is approximate for practical applications due to the finite length of the scanning trajectory. In a multi-linear scanning trajectory modes, the formula Eq.(6) can be modified to achieve theoretically exact results as[28]

$$f(\mathbf{r}) = \sum_{i=1}^{N} \int_{\lambda_b}^{\lambda_e} \frac{d^2}{U^2} \int_{t_{\min}}^{t_{\max}} \frac{h\sec(\lambda+\psi_i)w(\psi_i,\lambda,\mathbf{n})}{\sqrt{d^2 + 2t_i d \sin(\lambda+\psi_i) + t_i^2}} p(t_i,\lambda,\psi_i) g(t_i'-t_i) d\lambda dt_i, \tag{8}$$

where $w(\psi_i,\lambda,\mathbf{n})$ is a normalized weighting function which will be discussed in following section, $N$ is the number of straight segments of a multiple translation scanning trajectory mode, $t_i'$ is the local coordinate for the x-ray passing through the point $\mathbf{r}$ and its value can be determined by $t_i' = d\left(\frac{\mathbf{e}_i \cdot \mathbf{d}_{2_i}}{\mathbf{e}_i \cdot \mathbf{d}_{1_i}} - \tan(\lambda+\psi_i)\right)$, $\mathbf{e}_i$ is the unit vector starting from $\boldsymbol{\Phi}(\psi_i,\lambda)$ to $\mathbf{r}' \in R^2$ on the object image, and two unit vectors are defined as $\mathbf{d}_{1_i} = (-\sin\psi_i, \cos\psi_i)$ and $\mathbf{d}_{2_i} = (\cos\psi_i, \sin\psi_i)$. This FBP-type algorithm is similar to the results reported by Kong and Yu [2]. In this paper, we only consider a single translation (1T), two orthogonal translations (2T) and three symmetric translations (3T) for PTCT system (Fig. 2).

## 3. BPF-type PTCT Algorithms

### 3.A. Backprojection Step

According to the classic BPF algorithm, image can be reconstructed chord-by-chord where two ends of each chord are on the scanning trajectory. However, this concept should be modified for PTCT because all chords overlap on the same line-segment. Similar to the classic BPF algorithm for generalized scanning trajectories, one can consider L-PI lines and the intersections of the object function with the L-PI lines are support segments.

The backprojection step of the BPF algorithm yields an intermediate Hilbert image function $b_{(\psi,\lambda_b,\lambda_e)}(\mathbf{r})$ on the L-PI line

$$b_{(\psi,\lambda_b,\lambda_e)}(\mathbf{r}) = \int_{\lambda_b}^{\lambda_e} \frac{1}{|\mathbf{r}-\mathbf{\Phi}(\psi,\lambda)|} \frac{\partial}{\partial q} p(t,q,\psi)\Big|_{q=\lambda,\mathbf{e}\text{ is fixed}} d\lambda \quad . \tag{9}$$

In order to simplify Eq. (9), the derivative of projection data $G(t,\lambda,\psi)$ can be written as

$$G(t,\lambda,\psi) = \frac{\partial}{\partial q} p(t,q,\psi)\Big|_{q=\lambda,\mathbf{e}\text{ is fixed}} = \left(\frac{\partial p}{\partial \lambda} + \frac{\partial p}{\partial t}\frac{\partial t}{\partial \lambda} + \frac{\partial p}{\partial \psi}\frac{\partial \psi}{\partial \lambda}\right)\Big|_{\mathbf{e}\text{ is fixed}} . \tag{10}$$

Based on Eq. (5), one can obtain

$$\begin{aligned}
\frac{\partial t}{\partial \lambda}\Big|_{\mathbf{e}\text{ is fixed}} &= d \frac{\partial\left(\frac{\mathbf{e}\cdot\mathbf{d_2}}{\mathbf{e}\cdot\mathbf{d_1}} - \tan(\lambda+\psi)\right)}{\partial \lambda} \\
&= d \frac{\partial\left(\frac{\mathbf{e}\cdot\mathbf{d_2}}{\mathbf{e}\cdot\mathbf{d_1}}\right)}{\partial \lambda} - d\sec^2(\lambda+\psi) \\
&= d \frac{(\mathbf{e}\cdot\mathbf{d_2})'\times(\mathbf{e}\cdot\mathbf{d_1}) - \mathbf{e}\cdot\mathbf{d_2}\times(\mathbf{e}\cdot\mathbf{d_1})'}{(\mathbf{e}\cdot\mathbf{d_1})^2} - d\sec^2(\lambda+\psi) \\
&= -d\sec^2(\lambda+\psi)
\end{aligned} \tag{11}$$

$$\frac{\partial \psi}{\partial \lambda} = 0 \quad . \tag{12}$$

In Eq. (11) the relationships $\mathbf{d_1}'=0$, $\mathbf{d_2}'=0$ have been used in the above derivation. Substituting Eqs. (11) and (12) into Eq. (10), $G(t,\lambda,\psi)$ can be re-expressed as

$$G(t,\lambda,\psi) = \frac{\partial p}{\partial \lambda} - d\sec^2(\lambda+\psi)\frac{\partial p}{\partial t}. \tag{13}$$

Substituting Eqs. (1) and (13) into Eq. (9), and replacing **r** with $(x,y)$, finally, it yields

$$b_{(\psi,\lambda_b,\lambda_e)}(x,y) = \int_{\lambda_b}^{\lambda_e} \frac{\left(\frac{\partial p}{\partial \lambda} - d\sec^2(\lambda+\psi)\frac{\partial p}{\partial t}\right)}{\sqrt{\left(x+\frac{h}{\cos(\psi+\lambda)}\sin\lambda\right)^2 + \left(y+\frac{h}{\cos(\psi+\lambda)}\cos\lambda\right)^2}} d\lambda. \tag{14}$$

Noting the scanning trajectory cannot extend from minus infinity to infinity in 1T mode. We cannot obtain the theoretically exact Hilbert image along L-PI lines. In practical, we have to employ multiple finite translation scanning trajectories to guarantee theoretically exact reconstruction with an additional weighting for data redundancy.

Let $b(x,y)$ be theoretically exact Hilbert images from a multi-linear scanning trajectory modes, $b(x,y)$ can be expressed as follow

$$b(x,y) = \sum_{i=1}^{N} \int_{\lambda_b}^{\lambda_e} \frac{\text{sgn}(\mathbf{e_i}\cdot\mathbf{\chi}_\perp)\frac{\partial}{\partial q}\{w(\psi_i,\lambda,\mathbf{n})p(t_i,q,\psi_i)\}\big|_{q=\lambda, \mathbf{e_i} \text{ is fixed}}}{\sqrt{\left(x+\frac{h}{\cos(\psi_i+\lambda)}\sin\lambda\right)^2 + \left(y+\frac{h}{\cos(\psi_i+\lambda)}\cos\lambda\right)^2}} d\lambda, \tag{15}$$

where $w(\psi_i,\lambda,\mathbf{n})$ is a normalized weight function to explore redundant information contained in multiple-linear scanning trajectories modes, $\mathbf{e_i}$ is the unit vector starting from $\mathbf{\Phi}(\psi_i,\lambda)$ to $(x,y)\in R^2$ on the object image. $\mathbf{\chi}_\perp$ indicates a direction that is perpendicularly to the L-PI line.

**3.B. Finite Inverse Hilbert transform**

Let $f(x)$ be a 1D smooth function on a finite support $[x_b, \mathrm{x}_e]$. Let $h(x)$ be its Hilbert transform, then a Hilbert transform pair can be written as

$$\begin{cases} f(x) = pv \int_{x_b}^{x_e} \dfrac{h(x')}{\pi(x'-x)} dx' \\ h(x) = pv \int_{-\infty}^{+\infty} \dfrac{f(x')}{\pi(x-x')} dx' \end{cases}, \qquad (16)$$

where $pv$ represents the Cauchy principal value of the integral. Because the object is a finite support function on $[x_b, x_e]$, the true image can be reconstructed using the finite inverse Hilbert transform formula. The constant C [4, 8, 9, 24] plays an important role in reconstructing the image on the support segment, which cannot be obtained directly from the projections of PTCT. To tackle this thorny problem, we mainly adopt two finite inverse Hilbert formulae reported in [27] and [25]. The finite inverse Hilbert formula in [27] can be written as

$$f(x) = \dfrac{1}{2\pi} \sqrt{\dfrac{x-x_b}{x_e-x}} \int_{x_b}^{x_e} \sqrt{\dfrac{x_e-x}{x-x_b}} \dfrac{h(x')}{(x-x')} dx', \qquad (17)$$

and that in [25] can be read as

$$f(\psi, x) = \dfrac{1}{2\pi} \dfrac{1}{\sqrt{L^2-x^2} - \sqrt{l^2-x^2}} \int_{-L}^{L} \dfrac{k(L,l,\eta) h(x')}{(x-x')} d\eta, \qquad (18)$$

where $L > l \geq \max(|x_e|, |x_b|)$, $k(L, l, x)$ can be expressed as

$$k(L, l, x) = \begin{cases} -\sqrt{L^2-x^2} & l \leq |x| \leq L \\ \sqrt{l^2-x^2} - \sqrt{L^2-x^2} & |x| \leq l \\ 0 & \text{otherwise} \end{cases}. \qquad (19)$$

In formula (19), the true image is sensitive to the parameters $L$ and $l$. In practical, we can select an appropriate value over a range of $l = \max(|x_b|, |x_e|) + (2 \sim 3\, pixels)$, $L = (1.1 \sim 1.3) \max(|x_e|, |x_b|)$.

Based on the results obtained in section 3.A, $f(\mathbf{r})$ can be reconstructed from a multi-segment linear scanning mode

$$f(\mathbf{r}) = \sum_{i=1}^{N} \frac{1}{2\pi} \sqrt{\frac{x-x_b}{x_e-x}} \int_{x_b}^{x_e} \sqrt{\frac{x_e-x}{x-x_b}} \frac{b_{(\psi_i,\lambda_b,\lambda_e)}(x')}{x-x'} dx', \qquad (20)$$

and

$$f(\mathbf{r}) = \sum_{i=1}^{N} \frac{1}{2\pi} \frac{1}{\sqrt{L^2-x^2} - \sqrt{l^2-x^2}} \int_{-L}^{L} \frac{k(L,l,x') b_{(\psi_i,\lambda_b,\lambda_e)}(x')}{x-x'} dx'. \qquad (21)$$

For convenience, the above two formulae (20) and (21) can be viewed as two BPF-type algorithms, named MP-BPF and MZ-BPF algorithms, respectively.

**3.C. Weighted function for data redundancy**

For the PTCT acquisition modes, when only one straight-segment scanning trajectory is adopted, some data will be missed at some directions due to the finite length of scanning trajectory and it will cause severe limited-angle artifacts at the ends of angular range [29] (Fig.3).

To reconstruct theoretically exact images, multiple linear translations scanning modes was considered in [1, 2, 28]. However, it can introduce redundancy due to the fact that one projection along certain x-ray path may be measured two or multiple times. In order to explain it clearer, various quantities are defined (Fig.4). First, $\boldsymbol{\Phi}'(\psi,\lambda)\big|_{\psi \, fixed}$ is a vector that tangent to straight scanning path. It can be expressed as

$$\boldsymbol{\Phi}'(\psi,\lambda)\big|_{\psi \, fixed} = \left( \frac{-h\cos\psi}{\cos^2(\lambda+\psi)}, \frac{-h\sin\psi}{\cos^2(\lambda+\psi)} \right)\bigg|_{\psi \, fixed}, \quad \text{s.t } \lambda \in \left( -\frac{\pi}{2}-\psi, \frac{\pi}{2}-\psi \right). \qquad (22)$$

$\boldsymbol{\eta}$ is a unit vector indicating the direction of an x-ray path from the source $\boldsymbol{\Phi}(\psi,\lambda)$ to point $\mathbf{r}$ in the object support, which can be defined as

$$\boldsymbol{\eta} = \frac{\mathbf{r}-\boldsymbol{\Phi}(\psi,\lambda)}{\|\mathbf{r}-\boldsymbol{\Phi}(\psi,\lambda)\|}. \qquad (23)$$

$\mathbf{m}(\psi,\lambda,\mathbf{r})$ is a unit vector that is orthogonal to $\mathbf{\eta}$ with sign chosen so that $\mathbf{\Phi}'(\psi,\lambda)\cdot\mathbf{m}(\psi,\lambda,\mathbf{r})\geq 0$. Mathematically,

$$\mathbf{m}(\psi,\lambda,\mathbf{r}) = \frac{\mathbf{\Phi}'(\psi,\lambda)-(\mathbf{\Phi}'(\psi,\lambda)\cdot\mathbf{\eta})\mathbf{\eta}}{\|\mathbf{\Phi}'(\psi,\lambda)-(\mathbf{\Phi}'(\psi,\lambda)\cdot\mathbf{\eta})\mathbf{\eta}\|}. \tag{24}$$

Because $\mathbf{\eta}$ and $\mathbf{\Phi}'(\psi,\lambda)$ are not collinear, $\|\mathbf{\Phi}'(\psi,\lambda)-(\mathbf{\Phi}'(\psi,\lambda)\cdot\mathbf{\eta})\mathbf{\eta}\|$ is always greater than zero and $\mathbf{\Phi}'(\psi,\lambda)\cdot\mathbf{m}(\psi,\lambda,\mathbf{r})$ is never equal to zero.

Now, the definition of $w(\psi,\lambda,\mathbf{n})$ is given. Physically, $w(\psi,\lambda,\mathbf{n})$ is a weight to calculate the information redundancy in the dataset from a multi-linear translational scanning mode. According to formulae (3) and (4), fan-beam projections can furnish duplicated measurements for the reconstruction of $f(\mathbf{r})$ when their zeniths form a straight line containing $\mathbf{r}$. Indeed, for each zenith on the line orthogonal to a given direction $\mathbf{n}$, $\mathbf{\Phi}(\psi,\lambda)\cdot\mathbf{n}=\mathbf{r}\cdot\mathbf{n}$. It means the x-ray emitted from different zenith can produce same projection. The $w(\psi,\lambda,\mathbf{n})$ is designed to calculate this redundancy. Let $M(\psi,\lambda,\mathbf{n})$ be the number of intersections of the line $(\mathbf{r},\mathbf{n})$ with all linear paths, where $1\leq M(\psi,\lambda,\mathbf{n})\leq N$. For any fixed line passing through the object, the intersections with scanning trajectories are labelled as $\mathbf{\Phi}(\psi_i,\lambda)$, where $i=1,...,M(\psi,\lambda,\mathbf{n})$. Therefore, the normalized weight function can be defined as

$$\sum_{i=1}^{M(\psi,\lambda,\mathbf{n})} w_{(\mathbf{r},\mathbf{n})}(\psi_i,\lambda,\mathbf{n}) = 1. \tag{25}$$

One simple way is to choose $w(\psi,\lambda,\mathbf{n})=1/M(\psi,\lambda,\mathbf{n})$ as a weight function. However, in order to avoid $M(\psi,\lambda,\mathbf{n})$ discontinuity to suppress streak artifacts, we follow the strategy proposed in [30]

$$w(\psi_i, \lambda, \mathbf{n}) = \frac{\varsigma(\psi_i, \lambda)}{\sum_{i=1}^{M(\psi,\lambda,\mathbf{n})} \varsigma_{(\mathbf{r},\mathbf{n})}(\psi_i, \lambda)} \quad \text{s.t } \lambda \in \left(-\frac{\pi}{2} - \psi_i, \frac{\pi}{2} - \psi_i\right) \quad (26)$$

where $\varsigma(\psi_i, \lambda)$ is a positive function which disappears at the terminal point of each .linear translation. In this paper, $\varsigma(\psi_i, \lambda)$ is selected as

$$\varsigma(\psi_i, \lambda) = \begin{cases} e^{\frac{(\lambda - \lambda_b - \varepsilon)^2}{(\lambda - \lambda_b - \varepsilon)^2 - \varepsilon^2}} & \lambda \in [\lambda_b, \lambda_b + \varepsilon) \\ 1 & \lambda \in (\lambda_b + \varepsilon, \lambda_e - \varepsilon) \\ e^{\frac{(\lambda - \lambda_e + \varepsilon)^2}{(\lambda - \lambda_e + \varepsilon)^2 - \varepsilon^2}} & \lambda \in (\lambda_e - \varepsilon, \lambda_e] \\ 0 & else \end{cases} \quad .. \quad (27)$$

the parameter $\varepsilon$ is a small positive value which play an important role in reducing the limited angle artifacts [29]. Note that the function of $\varsigma(\psi_i, \lambda)$ satisfies $\varsigma(\psi_i, \lambda = \lambda_b \text{ and } \lambda = \lambda_e) = 0$. It means the projection at the directions of $\lambda = \lambda_b$ and $\lambda = \lambda_e$ is not used by the algorithm. To make more use of the data, we further set $\varsigma(\psi_i, \lambda_b) = \delta$ and $\varsigma(\psi_i, \lambda = \lambda_b \text{ and } \lambda = \lambda_e) = \delta$, for some small $\delta$, implementation the strategy of reducing limited-angle artifacts in the practical.

**Example 3.1.** Some x-ray integrals are measured twice and the others are measured only once by using of 2T mode (Fig. 5). To verify the proposed weight function, we display the reconstructed images with and without weight function in Fig. 6.

To sum up, we obtain generalized BPF-type algorithms for image reconstruction from a series of linear translational scanning fan-beam data. Figure 7 is a flowchart summarizing the process of the developed algorithms.

**3.D. ROI-Image reconstruction in PTCT**

In the previous subsections, we only consider the case that all projections are nontrucated. However, it is common that the entire object cannot be covered by the field of view (FOV), that is, the projections are truncated. This can be demonstrated in Fig. 8. The reconstructed ROI images from FBP-type algorithms are with severe truncated artifacts from PTCT data. However, the BPF-type algorithms can accurately recover ROI images from truncated PTCT data under certain condition.

For an L-PI line specified by parameters $(\psi_i, \lambda_b, \lambda_e)$, as shown Eqs. (20) and (21), one only needs knowledge of the backprojection images in $[x_b, x_e]$ and $[-L, L]$ for ROI images reconstruction. Such knowledge cannot be obtained from 1T mode. However, it can be achieved by adding line-segment to the scanning locus when the support segment located in ROI can be fully illuminated, for example 2T and 3T modes. It is worth noting that the constant C is available by selecting appropriate L-PI line and hence the conventional finite inverse Hilbert transform can be used for 2T and 3T modes.

## 4. Numerical simulations

The proposed MP-BPF and MZ-BPF algorithms for fan-beam PTCT are extensively evaluated by using a modified Shepp-Logan phantom and an abdominal image phantom (Fig. 9). To characterize the performance of the proposed BPF-type algorithms in processing truncated projections compared with FBP-type algorithms, an ROI is selected as indicated by the circle in the abdominal image phantom, and the ROI radius is 42 mm. Both of the phantoms consist of $256 \times 256$ pixels each of which covers an area of $1mm \times 1mm$ mm$^2$. Other imaging parameters for 1T, 2T and 3T are summarized in Table 1. We numerically generated nontruncated PTCT data assuming 1000 detector cells with 500, 1000 and 1500 views uniformly distributed over 1T, 2T, 3T scanning trajectories, respectively. To simulate truncated projections, we extracted the projections on the central 590 detector cells. Uniformly distributed Gaussian noise was also added to make projections realistic, and 0.37% of the maximum value of noise-free projections was chosen as standard variance. The proposed BPF-type algorithms and the FBP-type algorithm were applied to reconstruct full and ROI images from nontruncated and truncated data, respectively.

To investigate the relationship between the reconstructed ROI and detector length, it needs rigid geometrical analysis. We assume the detector length is sufficient large for covering the red rectangle (Fig.10. Let $S$ represents an endpoint of scanning trajectory, and $\chi_0$ indicates the direction of L-PI lines. The region covered by blue rectangular should be slightly larger than corresponding the compact support function, which can guarantee all $\chi_0$ out of the support segment. For convenience, let us define the intersections of the blue rectangle with the edges of image phantom as $A$, $B$, $C$ and $D$, respectively. The intersection of x-ray starting from source point $S$ and passing through the point $A$ with the detector array is $A'$. Similarly, the x-ray **SC** intersects with detector array at $C'$. We assume the radius of ROI is $R$. In addition, we further assume that compact supported phantom is limited in the rectangular section. In other words, we can suppose that abdominal image phantom consists of $\mu \times \mu$ mm². Therefore, the length of detector can be written as $L = A'O' + O'C'$. Note that symmetry of detector structure in the PTCT, $L$ can be modified $L = 2\max(A'O', O'C') = 2O'C'$. The coordinate system is established in

Fig. 10, and the equation of $SC$ can be expressed

$$y+R=\frac{-R+h}{\mu/2+h\tan\lambda}(x-\mu/2). \tag{28}$$

While the detector is fixed on $y=d-h$. The length of detector array $L$ can be simplified as

$$L=\frac{2d(\mu/2+R(\mu/2+h\tan\lambda))}{h(h-R)}. \tag{29}$$

We remark that the needed detector using this formula is approximate. Because in practical, the blue rectangle may be much greater than the corresponding support segment. Again, the detector array can be shorter than that in theory in this case. Based on this, substituting specified value into the formula (29), we can obtain $L=581.2mm<590mm$ in Table 1. This demonstrates the selected ROI can be covered by the central 590 detector cells. Obviously, the selected ROI can also be covered by central 590 detector cells in 2T mode and 3T mode (Fig. 11).

## 4.1 Image reconstruction from nontruncated data

Figure 12 shows reconstructed Shepp-Logan images from nontruncated noise-free fan-beam data. Fig. 13 shows representative profiles on line $y=0$ reconstructed by different algorithms for different modes. To further quantitatively evaluate the reconstructed image quality, the root mean square error (RMSE) is calculated as in Table. 2. Reconstructed Shepp-Logan images from nontruncated noisy data are displayed in Fig. 14.

From the above results, one can see that exact reconstructions can be obtained through both 3T and 2T scanning modes and the image quality of 3T is better than 2T

That may because the projection data collected by 3T mode is more efficient comparable with 2T mode. The fact is that, 2T scanning mode, which can be regarded as a typical short scan, is difficult to determine the weight function in practically. However, the proposed weight function can deal with this redundancy as demonstrated in Fig. 6. In addition, the BPF-type algorithms in 3T mode are slightly inferior to the FBP-type algorithm with nontruncated projection data. This can be easily understood because an additional rebinning step is required to obtain the final images for BPF-type algorithms, which compromises the reconstructed image quality. Besides, it turned out that MP-BPF and MZ-BPF algorithms perform well for noisy projections.

## 4.2 ROI-Image Reconstruction from truncated data

The proposed BPF-type algorithms were applied to reconstruct the ROI in the abdominal image phantom in Figure 9. Fig. 15 shows reconstructed entire images from noise-free truncated projections. Compared with the results from FBP-type algorithm, reconstructed images from BPF-type algorithms are free of truncation artifacts. In addition, the BPF-type algorithms can be able to reconstruct almost the entire images beyond the ROI if the projections are truncated not too severe. To further investigate the performance of proposed BPF-type algorithms for ROI images reconstruction, Fig. 16 displays the ROI images extracted from Fig. 15, and Fig. 17 provides the profiles along the line $y=0$ in the images in Fig. 16. To further quantitatively evaluate the reconstructed ROI images, RMSEs were computed in Table 3.

In Fig. 17 (left column), one can observe truncation artifacts from the results by using FBP-type algorithm. In contrast, accurate ROI images can be obtained using BPF-type algorithms, and they are free of truncation artifacts. In 1T mode, we cannot accurately reconstruct ROI images by neither FBP-type algorithm nor BPF-type algorithms due to incompleteness of projections. However, we can obtain accurate ROI images in 3T than 2T modes using BPF-type algorithms as shown in Table 3.

## 5. Discussion and Conclusion

Based on the classic BPF-type algorithms for general scanning trajectories, we developed two algorithms (MP-BPF and MZ-BPF) on L-PI lines for image reconstruction from projections collected by different linear translational scanning modes (1T, 2T and 3T) in PTCT. Compared with the FBP-type algorithm [2], the proposed MP-BPF and MZ-BPF algorithms can capable of not only approximately recovering the entire object images under the condition that the truncation is not so severe, but also accurately reconstructing the ROI images from truncated PTCT data without truncation artifacts. In addition, we constructed a practical weight function to handle redundant projections from multi-linear translational scanning modes. Our experiments verified that the proposed algorithms can significantly improve the reconstructed image quality.

As illustrated in Fig. 13, from the truncated PTCT projections, we can only approximately obtain entire images although it has a great potential in accurately reconstructing a whole image using BPF-type algorithms. Obviously, it is of practical significance to further investigate the condition of reconstructing entire images from truncated PTCT data. The BPF-type algorithms can introduce image blurring when Hilbert images are rebinned to the grid in global coordinate system. As a result, this can reduce the reconstructed image quality compared with FBP-type algorithm for nontruncated projections as illustrated in Fig. 12.

Now, cone-beam computed tomography (CBCT) has been widely applied to guide, supervise and assess different imaging tasks in biomedical and non-biomedical fields [31-34]. It is a fascinating question on how to generalize our BPF-type algorithms of PTCT from fan-beam to cone-beam geometry. Generally speaking, it is impossible to develop theoretically exact reconstruction formula for off-plane image slices. However, it is possible to develop approximate reconstruction formulas using the concept of virtual L-PI lines [5]. More specifically, at first, the virtual L-PI lines are constructed to parallel to scanning trajectory in 3-D space. Second, a Hilbert image can be obtained from the weighted derivative of projection data. Finally, we can reconstruct images by performing finite inverse Hilbert filtering along the virtual L-PI lines.

In summary, we have developed the MP-BPF and MZ-BPF algorithms for fan-beam PTCT to exactly reconstruct ROI images. In the near future, we intend to extend them to cone-beam geometry for 3D PTCT reconstruction. We also plan to build an optimal experimental system to acquire real data and further optimize those algorithms for PTCT.

## Acknowledgments

This work is partly supported by National Natural Science Foundation of China (No.

61471070) and National Instrumentation Program of China (No. 2013YQ030629). The authors would like to thank Drs. Shaojie Tang, Xiaobin Zou and Xiaochuan Pan for valuable discussions and constructive suggestions.

Table 1 Parameters for numerical simulation

| Parameter | Value |
| --- | --- |
| Source to detector distance $d$ (mm) | 800 |
| Source to object distance $h$ (mm) | 600 |
| diameter of object (mm) | 181 |
| Detector array length for obtaining nontruncated data(mm) | 1000 |
| Detector array length for obtaining truncated data(mm) | 590 |
| Detector pixel size (mm) | 1 |
| Number of source points per translation $P$ | 500 |
| The distance of source per translation (mm) | 2078.5 |
| Translation mode | Equi-anglar |
| Reconstructed image size | 256×256 |
| Pixel size(mm$^2$) | 1x1 |

Table 2 RMSEs of reconstruction images from PTCT using different algorithms

| modified Shepp-Logan | 1T | 2T | 3T |
|---|---|---|---|
| FBP | 0.1301 | 0.0301 | 0.0199 |
| MZ-BPF | 0.1253 | 0.0305 | 0.0201 |
| MP-BPF | 0.1280 | 0.0322 | 0.0208 |

Table 3 RMSEs of ROIs images from PTCT using FBP-type and BPF-type algorithms

| abdominal phantom | 1T | 2T | 3T |
| --- | --- | --- | --- |
| FBP | 221.5463 | 547.28198 | 511.3447 |
| MP-BPF | 162.10961 | 23.40467 | 19.95901 |
| MZ-BPF | 182.44914 | 22.25198 | 19.65389 |

Fig. 1. General linear scanning trajectory of PTCT scanning mode.

Fig. 2. Geometrical illustration of the PTCT data acquisition modes. From left and right columns are for the 1T, 2T and 3T modes respectively

Fig. 3.Original image (left), FBP reconstruction (right) for an angular range, $(-\lambda, \lambda)$ with $\lambda=60°$ when $\psi=0$. Limited angle artifacts appear along the ends of the angular range with red solid lines in the reconstructed image.

Fig. 4. Definition of $\mathbf{m}(\psi,\lambda,\mathbf{r})$.

Fig. 5. 2T mode of PTCT. Some x-ray paths are measured twice and others are measured only once.

Fig. 6.Original image (left), FBP reconstruction without (middle) and with (right) weight function for an angular range, $\lambda=120°$ with a translation.

Fig. 7 Flowchart of generalized BPF algorithm for image reconstruction from fan-beam data collected from a series of line trajectories.

Fig. 8 Illustration of a small detector and ROI construction using BPF-type algorithms.

Fig. 9. A modified Sheep-Logan phantom (left) and an abdominal image phantom (right).

Fig. 10 Illustration of the selected ROI can be covered by 590 detector cells in 1T mode.

Fig. 11 Illustration of the selected ROI can be covered by 590 detector cells in 2T mode (1st row) and 3T mode (2st row).

Fig. 12 Reconstructed Shepp-Logan images from nontruncated noise-free fan-beam data by using FBP algorithm in (8) (1st column), the formula (15) (2nd column), the MP-BPF algorithm in (20) (3rd column), and the MZ-BPF algorithm in (21) (4th column), respectively. The first to third rows are from 1T, 2T and 3T modes, respectively. The display window is [0,1].

Fig. 13. Representative profiles on line $y=0$ by utilizing different algorithms and different modes. From left to right, the plots correspond to 1T, 2T and 3T modes, respectively.

Fig. 14. Same as Figure 12 but from noisy data.

Fig. 15. Reconstructed abdominal images from noise-free truncated fan-beam data by using FBP algorithm in (8) (1st column), MP-BPF algorithm in (20) (2nd column), and the MZ-BPF algorithm in (21) (3rd column), respectively. The first to third rows are from 1T, 2T and 3T modes, respectively. The display window is [-1000 HU, 1000 HU].

Fig. 16. Same as Figure 15 but only for ROI images.

Fig. 17. Profiles along $y=0$ in images in Fig. 16. From left to right, the plots correspond to 1T, 2T and 3T modes, respectively.

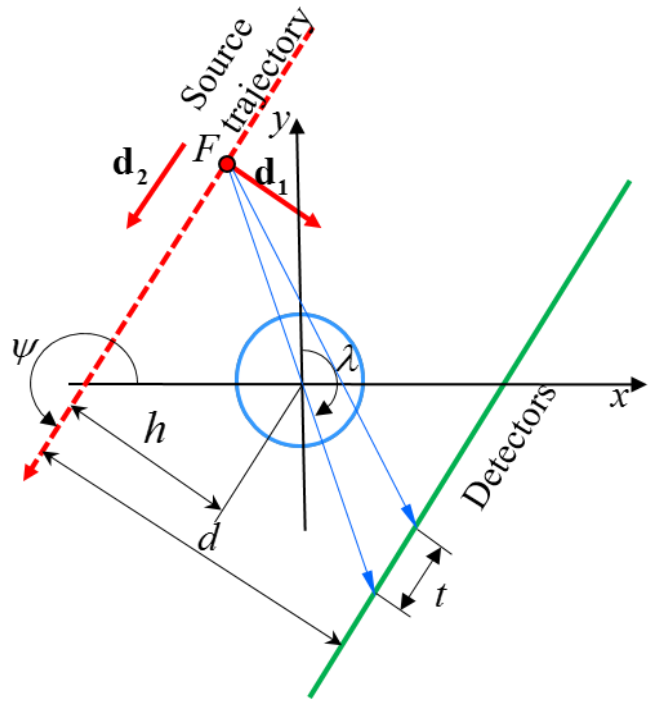

Fig. 1. General linear scanning trajectory of PTCT scanning mode.

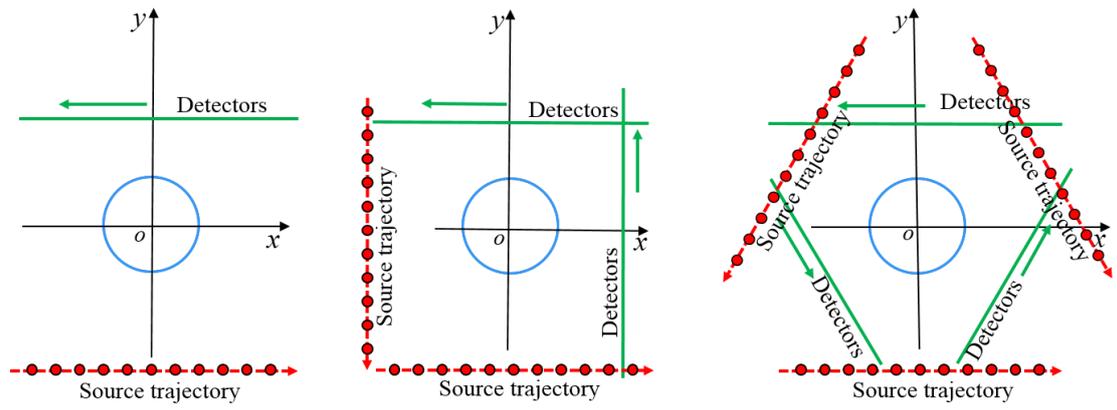

Fig. 2. Geometrical illustration of the PTCT data acquisition modes. From left and right columns are for the 1T, 2T and 3T modes respectively

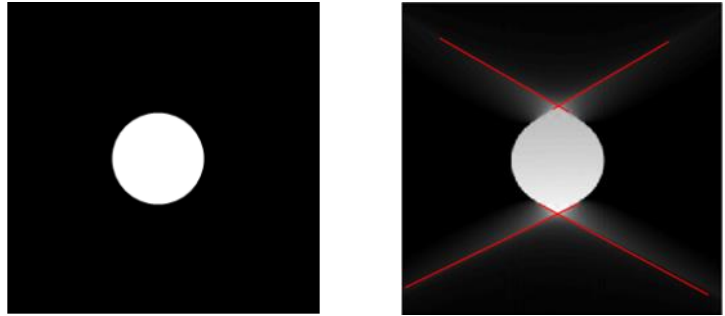

Fig. 3.Original image (left), FBP reconstruction (right) for an angular range, $(-\lambda,\ \lambda)$ with $\lambda=60°$ when $\psi=0$.

Limited angle artifacts appear along the ends of the angular range with red solid lines in the reconstructed image.

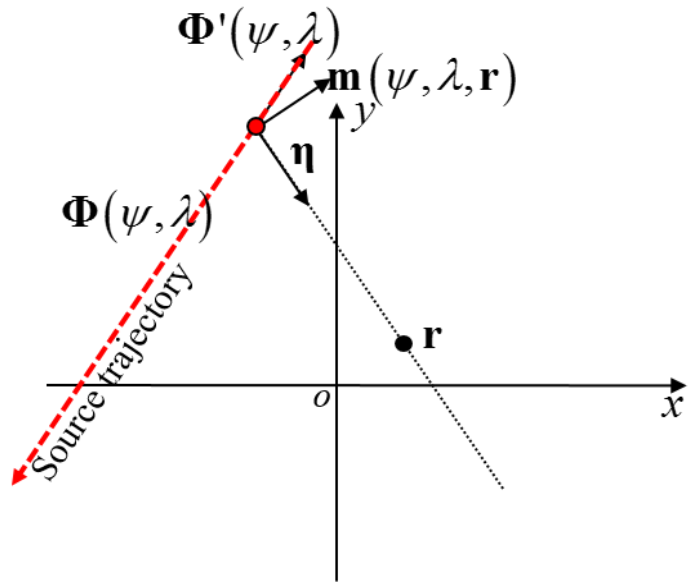

Fig. 4. Definition of $\mathbf{m}(\psi,\lambda,\mathbf{r})$.

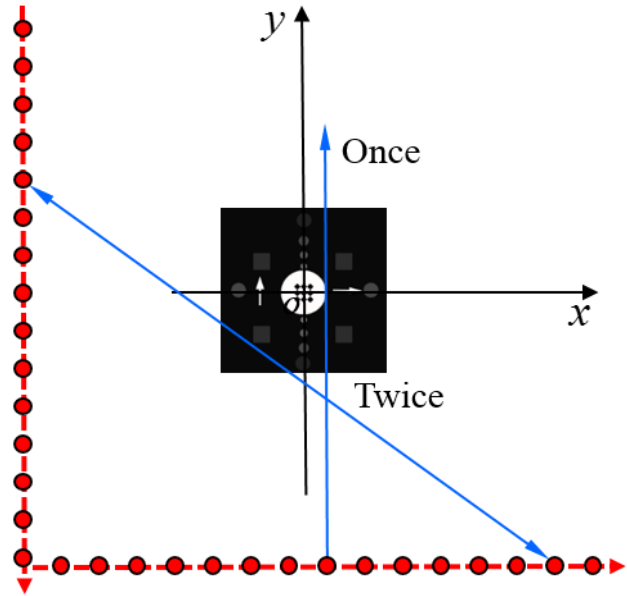

Fig. 5. 2T mode of PTCT. Some x-ray paths are measured twice and others are measured only once.

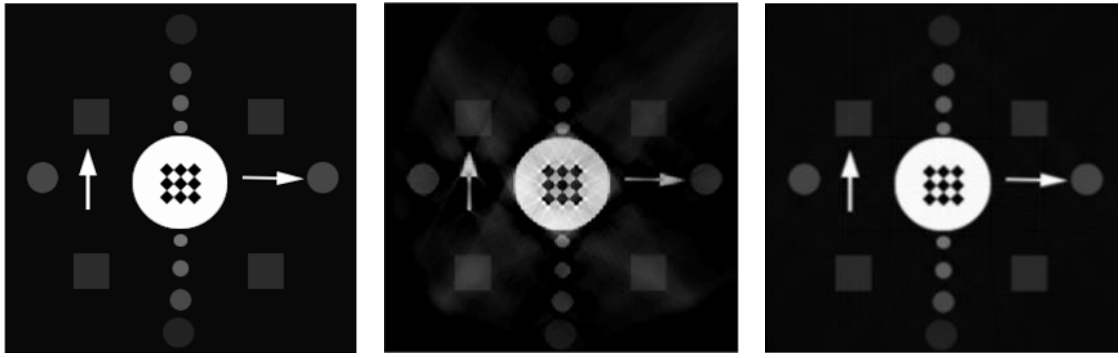

Fig. 6.Original image (left), FBP reconstruction without (middle) and with (right) weight function for an angular range, $\lambda=120°$ with a translation.

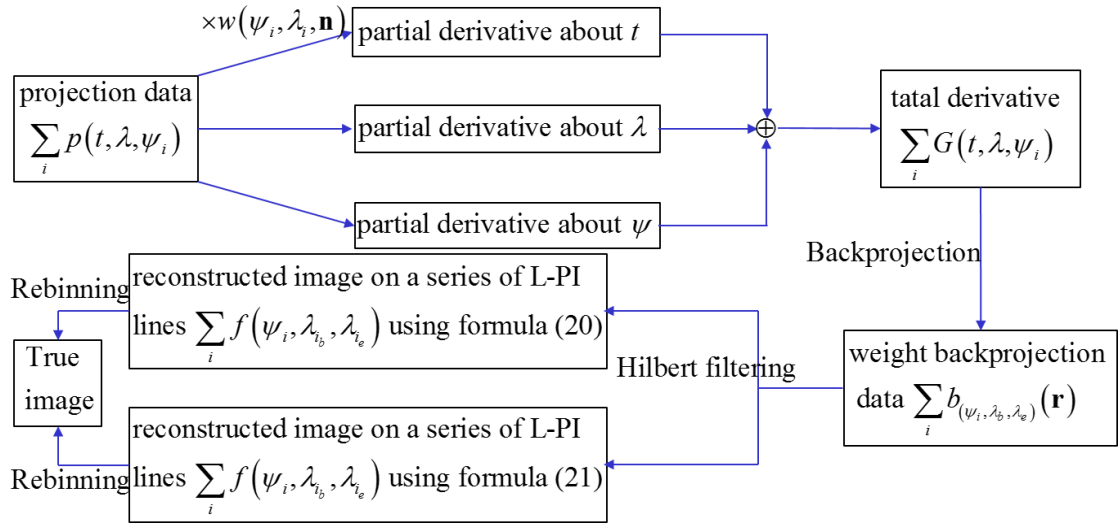

Fig. 7 Flowchart of generalized BPF algorithm for image reconstruction from fan-beam data collected from a series of line trajectories.

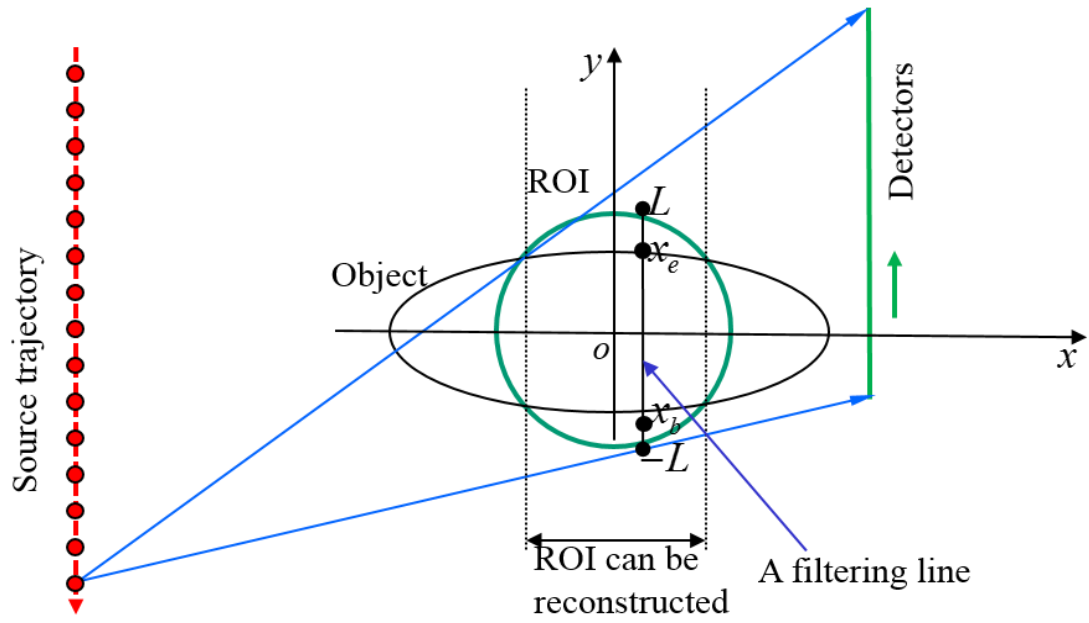

Fig. 8 Illustration of a small detector and ROI construction using BPF-type algorithms.

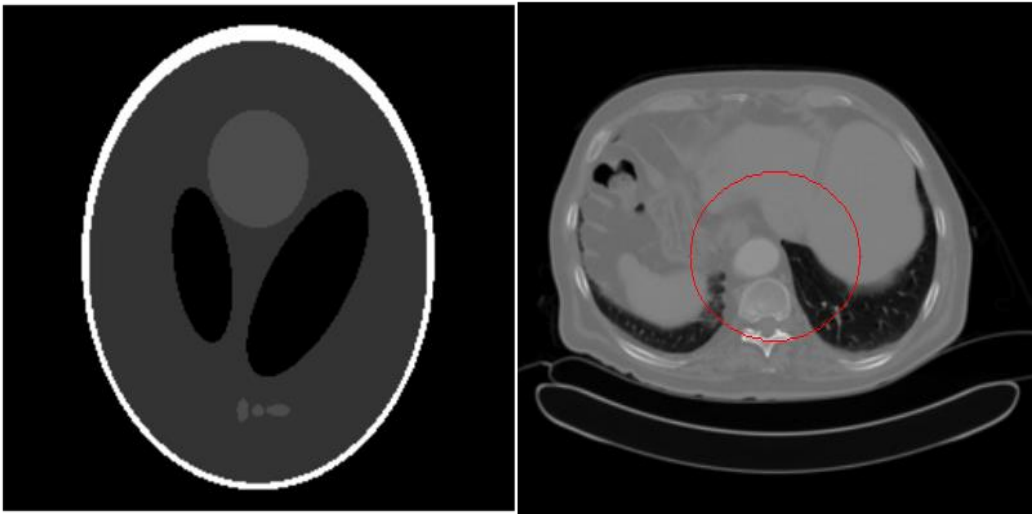

Fig. 9. A modified Sheep-Logan phantom (left) and an abdominal image phantom (right).

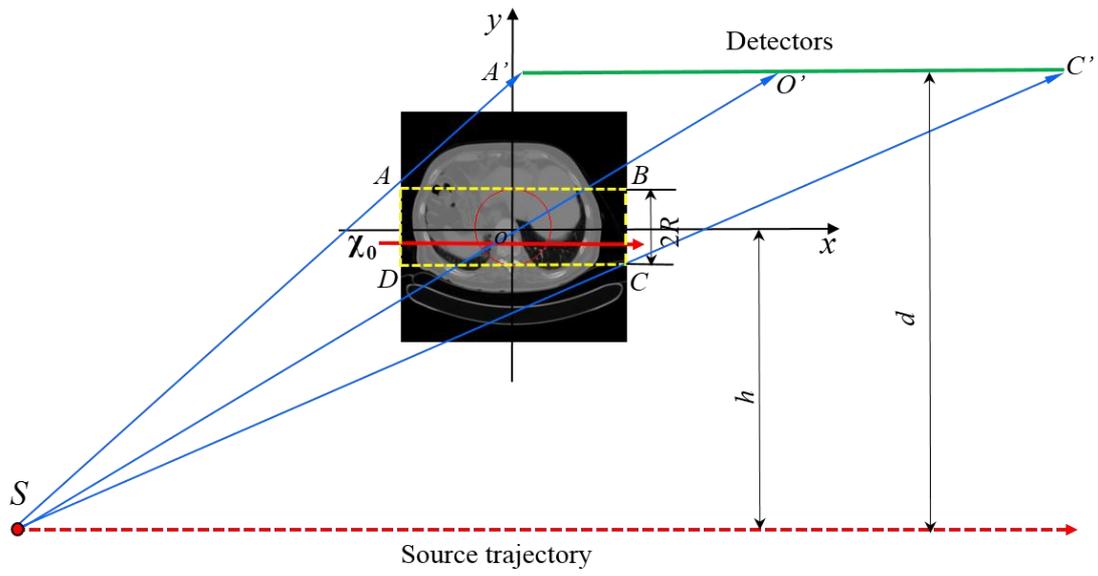

Fig. 10 Illustration of the selected ROI can be covered by 590 detector cells in 1T mode.

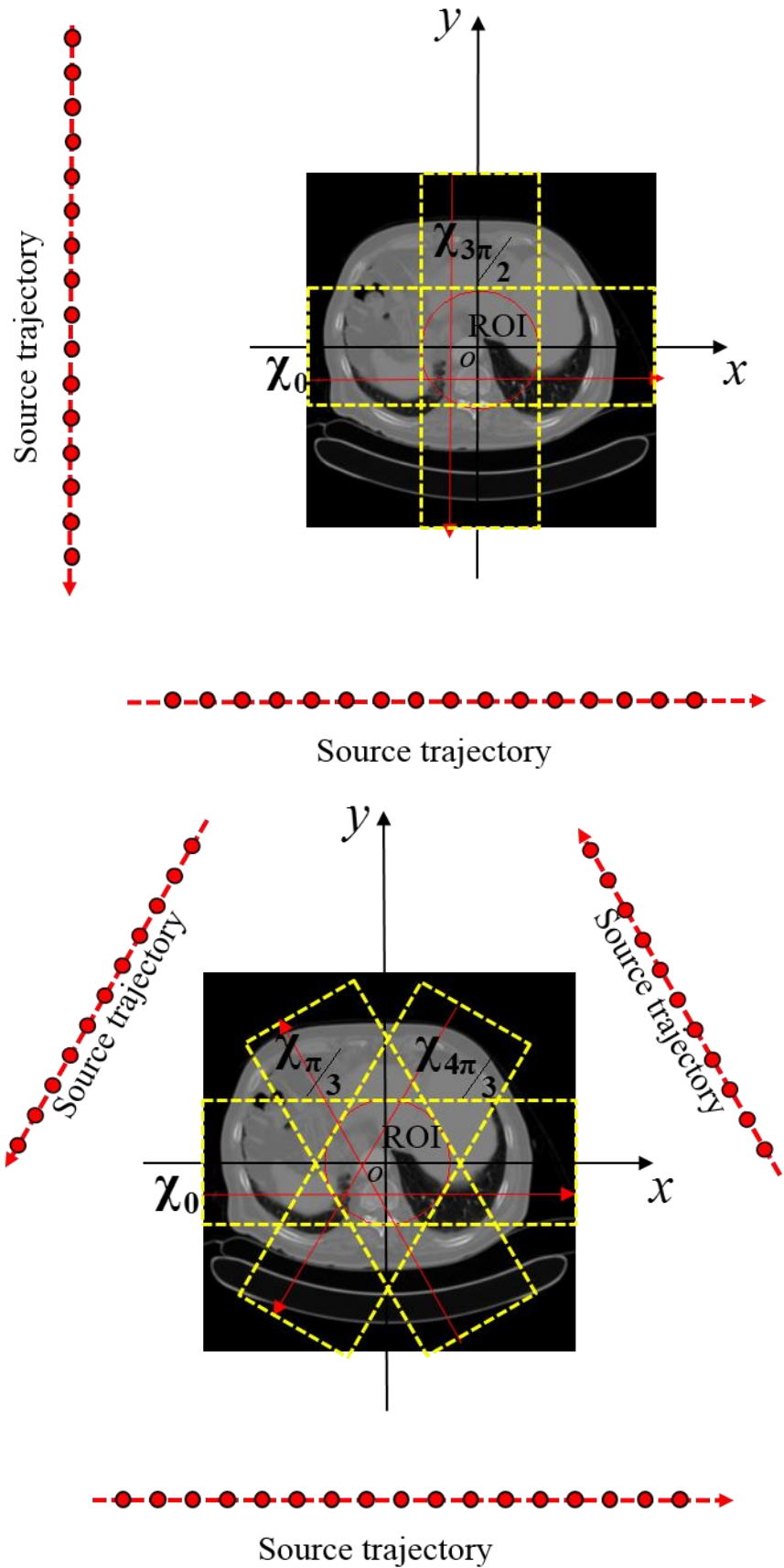

Fig. 11 Illustration of the selected ROI can be covered by 590 detector cells in 2T mode (1st row) and 3T mode (2st row).

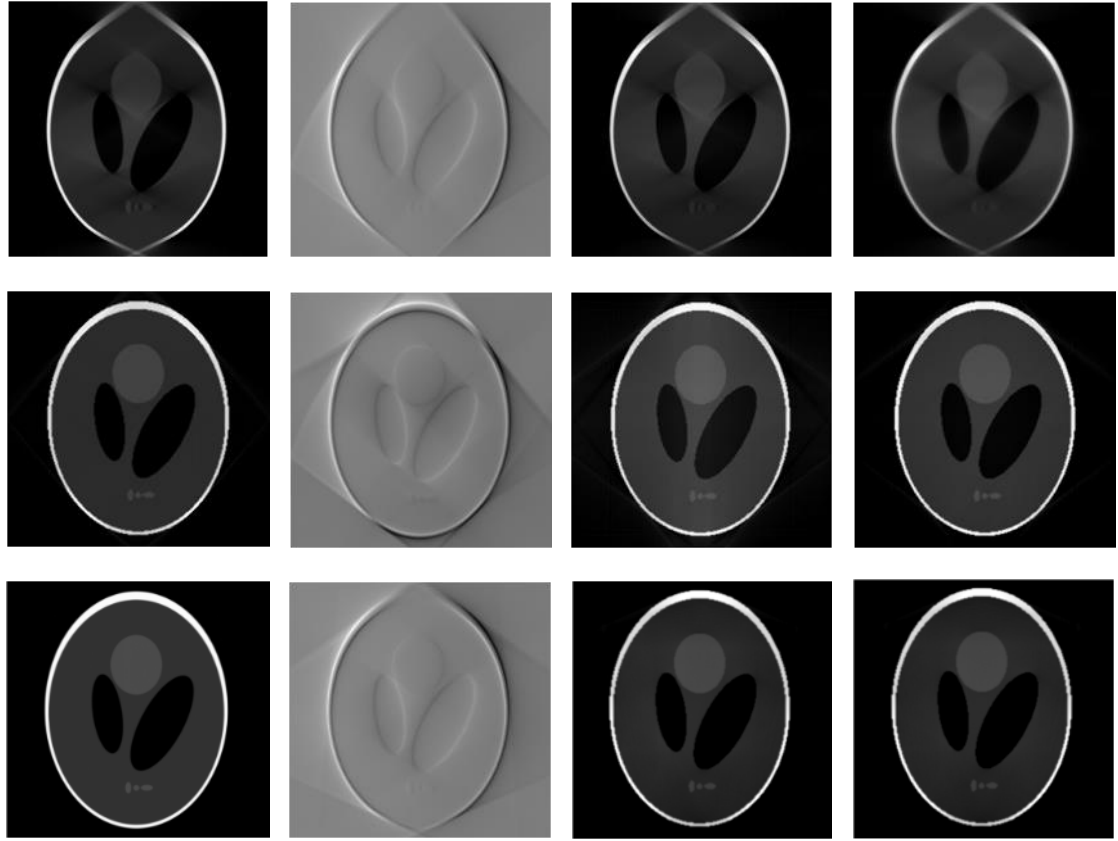

Fig. 12 Reconstructed Shepp-Logan images from nontruncated noise-free fan-beam data by using FBP algorithm in (8) (1$^{st}$ column), the formula (15) (2$^{nd}$ column), the MP-BPF algorithm in (20) (3$^{rd}$ column), and the MZ-BPF algorithm in (21) (4$^{th}$ column), respectively. The first to third rows are from 1T, 2T and 3T modes, respectively. The display window is [0,1].

*Corresponding author: Fenglin Liu, Key Lab of Optoelectronic Technology and Systems, Ministry of Education, Chongqing University, Chongqing 400044, China. Tel.: +86 23 8639 4828; E-mail: liufl@cqu.edu.cn

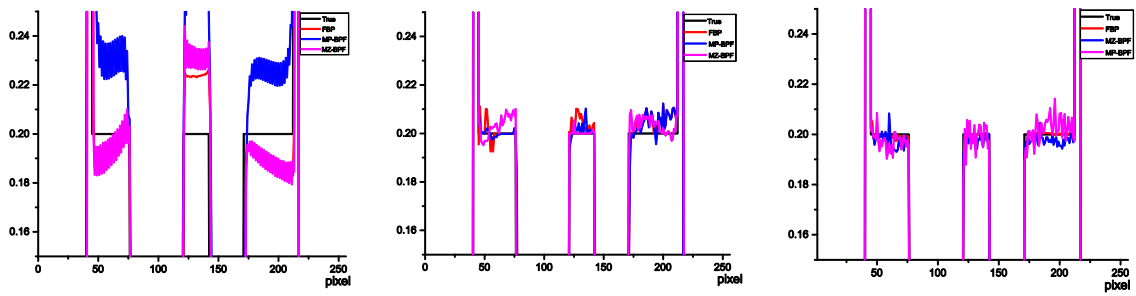

Fig. 13. Representative profiles on line $y = 0$ by utilizing different algorithms and different modes. From left to right, the plots correspond to 1T, 2T and 3T modes, respectively.

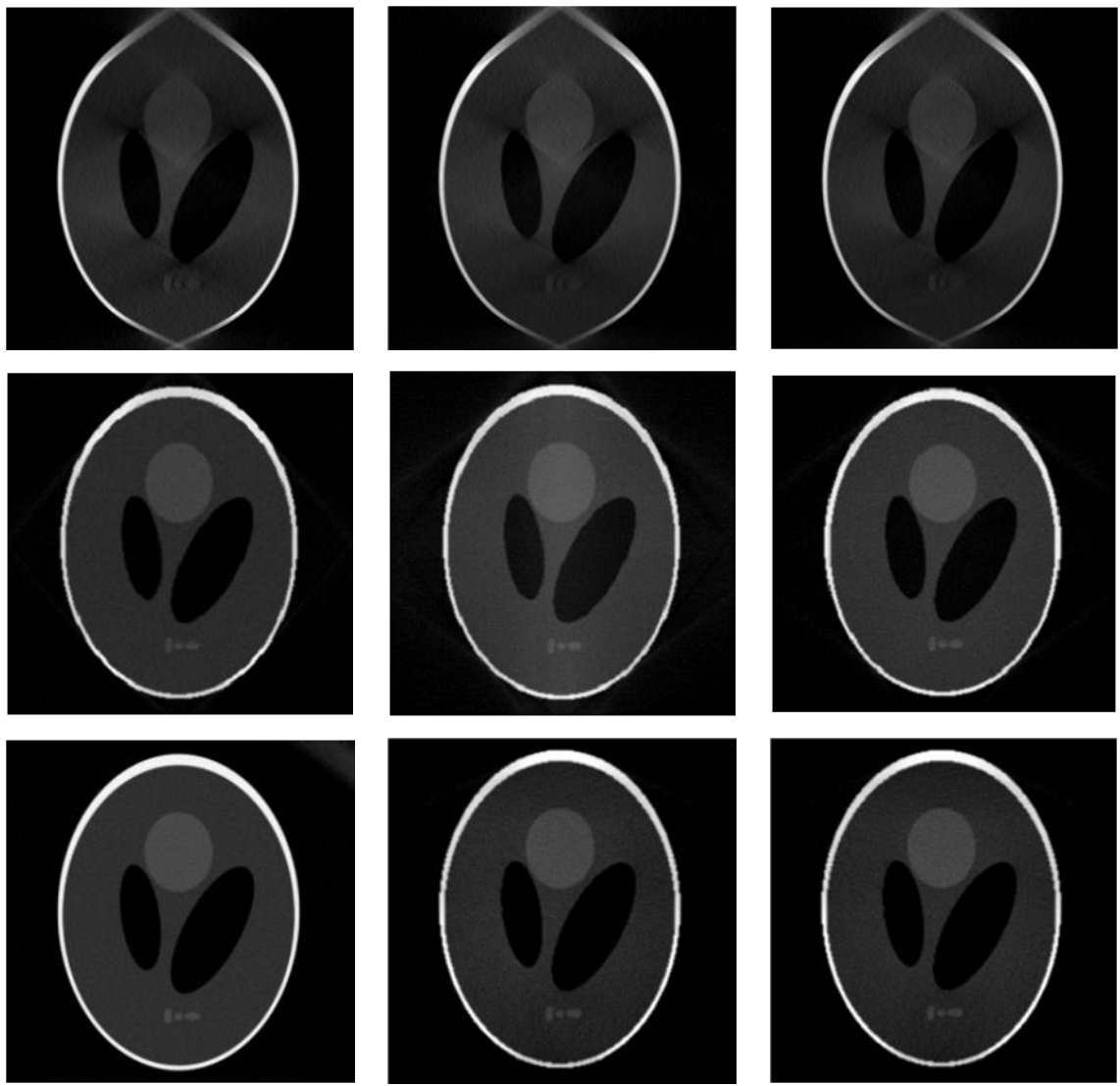

Fig. 14. Same as Figure 12 but from noisy data.

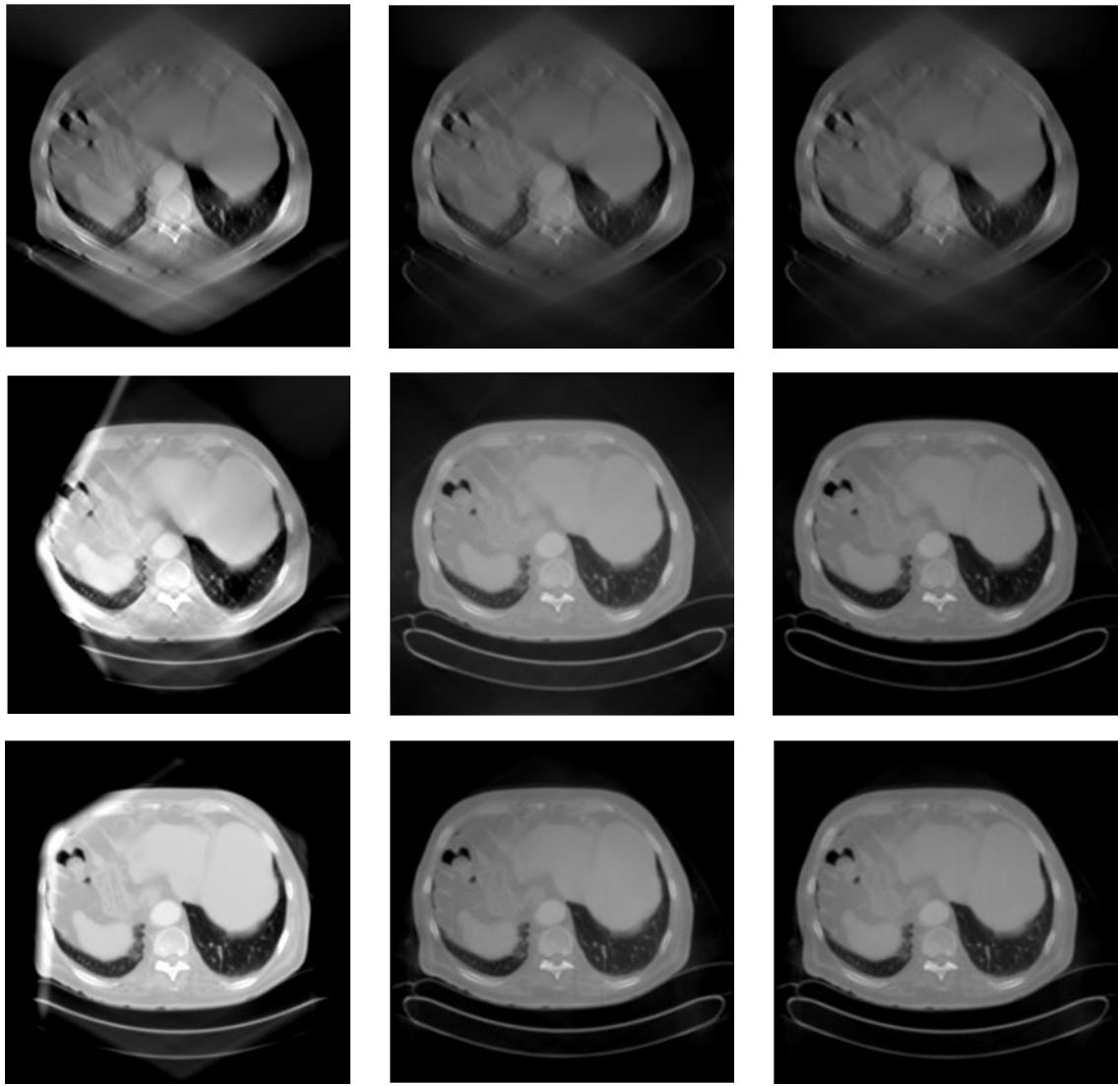

Fig. 15. Reconstructed abdominal images from noise-free truncated fan-beam data by using FBP algorithm in (8) (1$^{st}$ column), MP-BPF algorithm in (20) (2$^{nd}$ column), and the MZ-BPF algorithm in (21) (3$^{rd}$ column), respectively. The first to third rows are from 1T, 2T and 3T modes, respectively. The display window is [-1000 HU, 1000 HU].

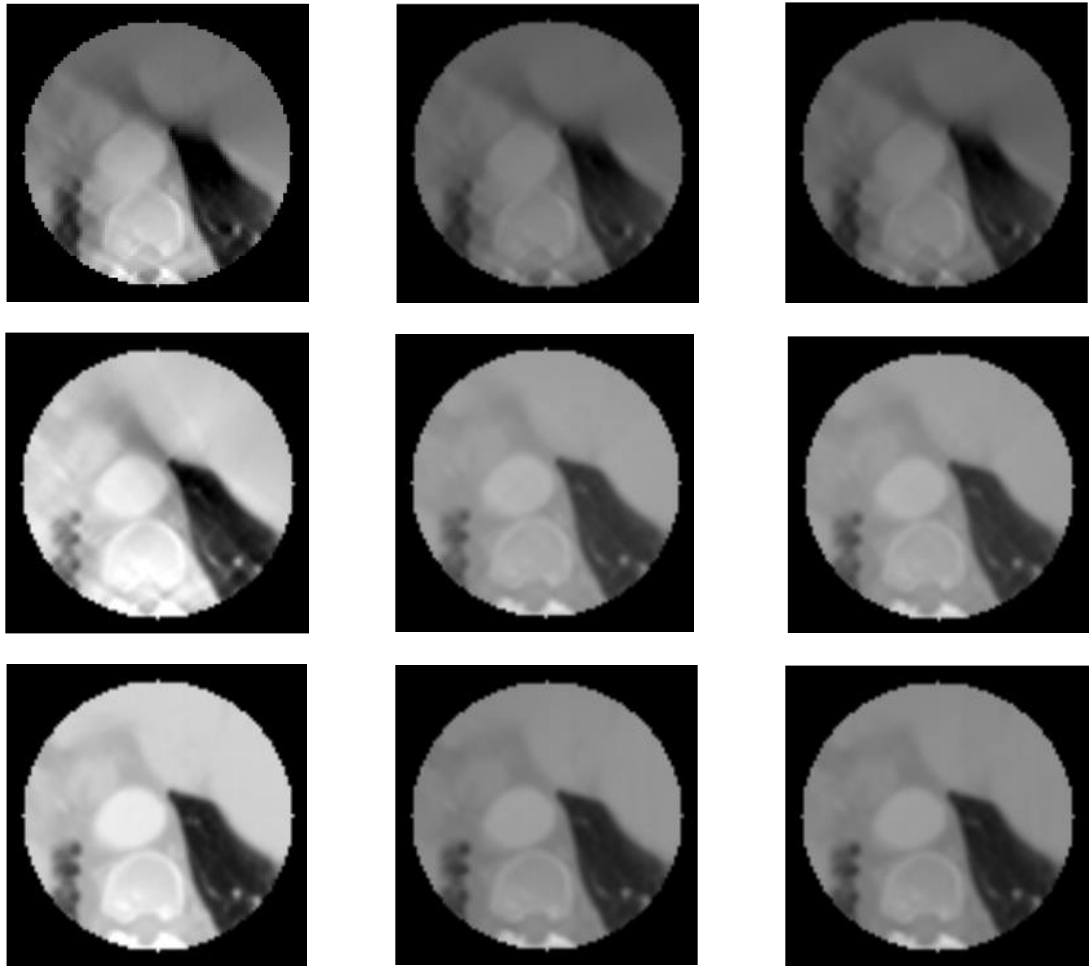

Fig. 16. Same as Figure 15 but only for ROI images.

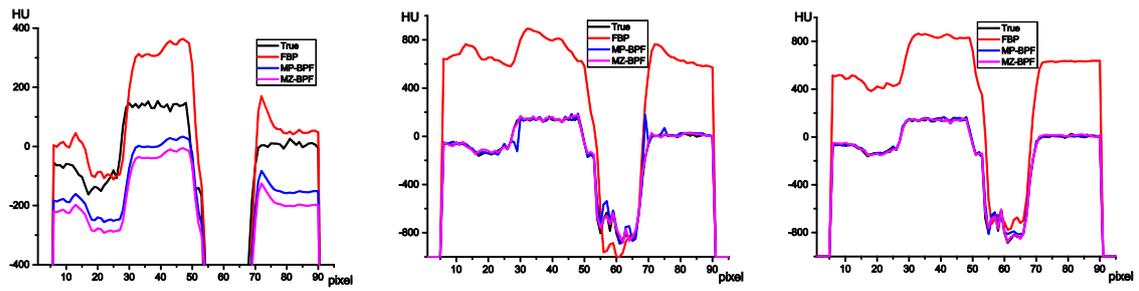

Fig. 17. Profiles along $y = 0$ in images in Fig. 16. From left to right, the plots correspond to 1T, 2T and 3T modes, respectively.